\documentclass[oldversion]{aa}
\def\folio{\ifnum\pageno=1\nopagenumbers\else\number\pageno\fi}

\def\lax    {\ifmmode{_<\atop^{\sim}}\else{${_<\atop^{\sim}}$}\fi}
\def\gax    {\ifmmode{_>\atop^{\sim}}\else{${_>\atop^{\sim}}$}\fi}
\newbox\grsign      \setbox\grsign=\hbox{$>$} 
\newdimen\grdimen   \grdimen=\ht\grsign
\newbox\simgreatbox \setbox\simgreatbox=\hbox{\raise.5ex\hbox{$>$}\llap
                        {\lower.5ex\hbox{$\sim$}}}\ht1=\grdimen\dp1=0pt
\newbox\simlessbox  \setbox\simlessbox =\hbox{\raise.5ex\hbox{$<$}\llap
                        {\lower.5ex\hbox{$\sim$}}}\ht2=\grdimen\dp2=0pt

\newbox\grsign \setbox\grsign=\hbox{$>$} \newdimen\grdimen \grdimen=\ht\grsign
\newbox\laxbox \newbox\gaxbox
\setbox\gaxbox=\hbox{\raise.5ex\hbox{$>$}\llap
     {\lower.5ex\hbox{$\sim$}}}\ht1=\grdimen\dp1=0pt
\setbox\laxbox=\hbox{\raise.5ex\hbox{$<$}\llap
     {\lower.5ex\hbox{$\sim$}}}\ht2=\grdimen\dp2=0pt
\def\gax{\mathrel{\copy\gaxbox}}
\def\lax{\mathrel{\copy\laxbox}}
\def\boxit#1    {\vbox{\hrule\hbox{\vrule\kern3pt
                  \vbox{\kern3pt#1\kern3pt}\kern3pt\vrule}\hrule}}
\def\h      {\ifmmode{^{\rm h}}\else{$^{\rm h}$}\fi}
\def\m      {\ifmmode{^{\rm m}}\else{$^{\rm m}$}\fi}
\def\s      {\ifmmode{^{\rm s}}\else{$^{\rm s}$}\fi}
\def\decas    {\ifmmode{{\rlap.}{''}}\else{${\rlap.}{''}$}\fi}
\def\mum     {\ifmmode{\mu{\rm m}}\else{$\mu{\rm m}$}\fi}
\def\s      {\ifmmode{^{\rm s}}\else{$^{\rm s}$}\fi}
\def\decdeg {\rlap . {}^\circ}     
\def\deg      {\ifmmode{^{\circ}}\else{$^{\circ}$}\fi}
\def\as     {\ifmmode {\rlap.}$\,$''$\,$\! \else ${\rlap.}$\,$''$\,$\!$\fi}
\def\decsec  {\ifmmode {\rlap.}$\,$^{s}$\,$\! \else ${\rlap.}$\,$^{s}$\,$\!$\fi}\def\decs  {\ifmmode {\rlap.}$\,$^{s}$\,$\! \else ${\rlap.}$\,$^{s}$\,$\!$\fi}

\def\kms    {\ifmmode{{\rm km~s}^{-1}}\else{km~s$^{-1}$}\fi}

\def\ccm    {cm$^{-3}$}

\def\Mspy   {\ifmmode {M_{\odot} {\rm yr}^{-1}} \else $M_{\odot}$~yr$^{-1}$\fi}
\def\Mdot   {\ifmmode {\dot M} \else $\dot M$\fi}
\def\mhd    {\ifmmode {n_{{\rm H}_2}} \else $n_{{\rm H}_2}$\fi}
\def\mhcd   {\ifmmode {N_{{\rm H}_2}} \else $N_{{\rm H}_2}$\fi}

\def\El      {\ifmmode{E_{\ell}}\else{$E_{\ell}$}\fi}
\def\beam    {\ifmmode{\theta_{\rm B}}\else{$\theta_{\rm B}$}\fi}
\def\mjyb   {\ifmmode {{\rm mJy~beam}^{-1}} \else{mJy~beam$^{-1}$}\fi}
\def\mujyb   {\ifmmode {\mu{\rm Jy~beam}^{-1}} \else{$\mu$Jy~beam$^{-1}$}\fi}
\def\Trot   {\ifmmode{T_{\rm rot}}\else$T_{\rm rot}$\fi}    
    
\def\Teff   {\ifmmode{T_{\rm eff}}\else$T_{\rm eff}$\fi}

\def\ITRS   {\ifmmode{\smallint {\rm T}_{R}^{*}dv}\else{$\smallint 
{\rm T}_{R}^{*}dv$}\fi}
\def\ITRS   {\ifmmode{\smallint {\rm T}_{R}^{*}dv}\else{$\smallint 
{\rm T}_{R}^{*}dv$}\fi}
\def\ITAS   {\ifmmode{\smallint {\rm T}_{A}^{*}dv}\else{$\smallint 
{\rm T}_{A}^{*}dv$}\fi}

\def\hzo        {H$_2$O}

\def\nhhh       {NH$_3$}

\def\lefttitle#1  {\noindent \hangindent=18.0pt \hangafter=1 {#1} \par}
\def\vol#1  {{\bf {#1}{\rm,}\ }}
\font\tenssb=cmssbx10
\font\tenbf=cmbx10
\font\sevenbf=cmbx8
\font\fivebf=cmbx6
\def\unetdemi    {\smallskipamount=6pt plus2pt minus2pt
                  \medskipamount=12pt plus4pt minus4pt
                  \bigskipamount=24pt plus8pt minus8pt
                  \normalbaselineskip=16pt plus0pt minus0pt
                  \normallineskip=2pt
                  \normallineskiplimit=0pt
                  \jot=6pt
                  {\def\smallskip {\vskip\smallskipamount}}
                  {\def\medskip   {\vskip\medskipamount}}
                  {\def\bigskip   {\vskip\bigskipamount}}
                  {\setbox\strutbox=\hbox{\vrule 
                    height17.0pt depth7.0pt width 0pt}}
                  \parskip 12.0pt
                  \normalbaselines}
\def\smallerspace {\smallskipamount=3pt plus0pt minus0pt
                  \medskipamount=6pt plus0pt minus0pt
                  \bigskipamount=10.5pt plus0pt minus0pt
                  \normalbaselineskip=10.5pt plus0pt minus0pt
                  \normallineskip=1pt
                  \normallineskiplimit=0pt
                  \jot=3pt
                  {\def\smallskip {\vskip\smallskipamount}}
                  {\def\medskip   {\vskip\medskipamount}}
                  {\def\bigskip   {\vskip\bigskipamount}}
                  {\setbox\strutbox=\hbox{\vrule 
                    height8.5pt depth3.5pt width 0pt}}
                  \parskip 0pt
                  \normalbaselines}
\def\memospace    {\smallskipamount=4pt plus1pt minus1pt
                  \medskipamount=6pt plus2pt minus2pt
                  \bigskipamount=14pt plus6pt minus6pt
                  \normalbaselineskip=14pt plus0pt minus0pt
                  \normallineskip=1pt
                  \normallineskiplimit=0pt
                  \jot=4pt
                  {\def\smallskip {\vskip\smallskipamount}}
                  {\def\medskip   {\vskip\medskipamount}}
                  {\def\bigskip   {\vskip\bigskipamount}}
                  {\setbox\strutbox=\hbox{\vrule 
                    height17.0pt depth7.0pt width 0pt}}
                  \parskip 2.0pt
                  \normalbaselines}
\def\memowidespace    {\smallskipamount=5pt plus1pt minus1pt
                  \medskipamount=7.5pt plus2pt minus2pt
                  \bigskipamount=17.5pt plus6pt minus6pt
                  \normalbaselineskip=17.0pt plus0pt minus0pt
                  \normallineskip=1.25pt
                  \normallineskiplimit=0pt
                  \jot=5pt
                  {\def\smallskip {\vskip\smallskipamount}}
                  {\def\medskip   {\vskip\medskipamount}}
                  {\def\bigskip   {\vskip\bigskipamount}}
                  {\setbox\strutbox=\hbox{\vrule 
                    height21.25pt depth8.75pt width 0pt}}
                  \parskip 2.5pt
                  \normalbaselines}
\usepackage{graphicx}
\usepackage{natbib}
\usepackage{txfonts}
\usepackage{ulem}
      \def\new#1 {{\bf #1 }}
      \def\cut#1 {\sout{#1} }

\begin{document}

\title{Submillimeter water and ammonia absorption by the
peculiar $z \approx 0.89$ interstellar medium in the gravitational lens
of the PKS 1830$-$211 system}
\author{K. M. Menten
\inst{1}
\and
R. G\"usten
\inst{1}
\and
S. Leurini
\inst{2}
\and
S. Thorwirth
\inst{1}
\and
C. Henkel
\inst{1}
\and
B. Klein
\inst{1}
\and
C. L. Carilli
\inst{2}
\and
M. J. Reid
\inst{3}
}


\offprints{K. M. Menten}

\institute{Max-Planck-Institut f\"ur Radioastronomie,
Auf dem H\"ugel 69, D-53121 Bonn, Germany
\email{kmenten, rguesten, sthorwirth, chenkel, bklein@mpifr-bonn.mpg.de}
\and
ESO, Karl-Schwarzschild-Strasse 2, D-85748 Garching bei M\"unchen, Germany
\email{sleurini@eso.org}
\and
Harvard-Smithsonian Center for Astrophysics
60 Garden Street
Cambridge, MA 02138, USA
\email{reid@cfa.harvard.edu}
}

\date{Received / Accepted}
\titlerunning{Redshifted water and ammonia toward PKS~1830$-$211}

\authorrunning{Menten et al.}

\abstract
{Using the Atacama Pathfinder Experiment (APEX) telescope we have
detected the rotational ground-state transitions of ortho-ammonia and
ortho-water toward the redshift $\approx0.89$ absorbing galaxy
in the PKS 1830$-$211 gravitational lens system. We discuss our
observations in the context of recent space-borne data obtained for these
lines with the SWAS and Odin satellites toward Galactic sources. We find commonalities, but also
significant differences between the interstellar media in a galaxy at
intermediate redshift and in the Milky Way. Future high-quality
observations of the
ground-state ammonia transition in PKS 1830$-$211, together with inversion line data, will lead to
strong constraints on the variation of the proton to electron
mass ratio over the past 7.2 Gyr.}


\keywords{Cosmology: observations --- Gravitational Lensing --- Astrochemistry --- ISM: abundances --- ISM -- molecules}

\maketitle

\section{\label{intro}Introduction}
The gravitational lens system involving the bright compact radio and millimeter source PKS 1830$-$211 is remarkable for several reasons:  A distant quasi-stellar object (QSO) at a redshift, $z$, of 2.507 \citep{Lidman1999} is imaged into two strong, milliarcsecond (mas) size sources separated by $\approx970$ mas, corresponding to
7.3 kpc at $z = 0.89$\footnote{Throughout this paper we are using
a $\Lambda$ cosmology with $H_0 = 73$~km~s$^{-1}$~Mpc$^{-1}$, $\Omega_{\rm M} = 0.28$, and $\Omega_{\Lambda} = 0.72$ \citep{Spergel2007}. 1 arcsecond corresponds to 7.54 kpc.} \citep{Jin2003}. These components appear embedded in extended emission, part of which, at high resolution, presents an Einstein ring \citep{Subrahmanyan1990,Jauncey1991}.

Through an act of ``cosmic conspiracy'', the line of sight to the background QSO passes through or near several galaxies \citep{Meylan2005} and, placed at Galactic longitude $+12\decdeg2$ and latitude  $-5\decdeg7$, the bulge of the Milky Way, causing local neutral hydrogen (HI) absorption \citep{Subrahmanyan1992}. \citet{Courbin2002} give a  historical summary of this complex picture. Most importantly,
the line of sight passes through a spiral galaxy at a redshift of $\approx0.89$ \citep{Winn2002}, in which continuum emission from the QSO is absorbed in lines of atomic hydrogen and hydroxyl (OH) \citep{Chengalur1999,KoopmansdeBruyn2005} and transitions from a variety of molecules in the interstellar media of two different spiral arms \citep{WiklindCombes1996,WiklindCombes1998}. The much deeper absorption has its centroid at $z = 0.88582$ (zero velocity in the heliocentric frame) and is toward the south-western (SW) image \citep{Frye1997, Muller2006}, which is rendered undetectable at optical wavelengths by the dust mixed with the molecular material \citep{Winn2002}. The absorption appears to be caused by material in  a spiral arm of the lensing galaxy at projected distance of $\approx 0\as4$ from the galaxy's center (3 kpc), clearly defined in the $I$-image presented by these authors.
We note that in the Milky Way the bulk of the molecular gas is distributed between Galactocentric radii of 4 and 7 kpc \citep{Bronfman1988}.

\citet{WiklindCombes1998} detected a second, weaker absorption cloud shifted by $-147$ \kms\ ($z = 0.88489$) relative to the primary absorption and toward the NE image \citep{Muller2006}, whose location 
is displaced by $0\as59$ (4.4 kpc) from galaxy's 
center.

For completeness, we mention that there is another HI absorber at $z=0.19$ that absorbs the NE and, weaker, the SW image as well as part of PKS 1830$-$211's extended
emission \citep{Lovell1996}.  The nature of this low opacity ($\tau_{\rm HI} \sim 0.01$) system's host is uncertain but possibly we are dealing with another intervening spiral galaxy of low surface brightness. No molecular absorption has been detected toward it \citep{WiklindCombes1998}.

T. Wiklind and F. Combes detected absorption in CO and in a number of
molecules with (much) higher dipole moments, namely HCN, HCO$^+$,
CS, and N$_2$H$^+$ \citep{WiklindCombes1996, WiklindCombes1998}.
\citet{Menten1999} added C$_2$H, HC$_3$N, and C$_3$H$_2$ to this list.
A number of rare isotopologues of C-, N-, O-, and S-bearing species were
imaged by \citet{Muller2006}, who found isotopic ratios differing from
Milky Way values.
\citet{Menten1999} compared PKS 1830$-$211's molecular make-up  with
Galactic dark clouds and found many similarities.
Recently, \citet{Henkel2008} observed a total of
ten metastable inversion transitions of ammonia (NH$_3$) and found
the observations to be consistent with 80--90\%\ of the ammonia-bearing
gas being warm ($\sim 80$ K). Amazingly, to explain the intensities of
the lines at high energy levels (up to 1000 K above the ground-state)
they have to invoke a hot $>600$ K medium for which in the Milky Way the
only known counterpart would be the Sgr B2 region near the Galactic center \citep{Huettemeister1995, Wilson2006}.

Redshifted absorbers provide the possibility to detect spectral lines that are unobservable from the ground due to attenuation by the Earth's atmosphere, namely low excitation water (\hzo) lines that cause the bulk of the terrestrial absorption. This was exploited by \citet{CombesWiklind1997}, who discovered absorption in the $1_{10}-1_{01}$ ground-state transition of ortho-water (o-\hzo) toward the ``other'' rich molecular intermediate-redshift ($z=0.685$) system B0218+357. The terrestrial, pressure-broadened $1_{10}-1_{01}$ \hzo\ line near 557 GHz causes the broadest range of absorption in the Earth's atmosphere below 1 THz, rendering the $\pm50$ GHz around it practically impossible to observe from the ground, even from the excellent 5100 m high site of the APEX telescope. Considerable effort was put in the investigation of this line in the Milky Way; it was a main motivation for two space missions, the Submillimeter Wave Astronomy Satellite \citep[SWAS][]{Melnick2000} and Odin \citep{Nordh2003,Hjalmarson2003}. Among many other things, Odin also delivered the first astronomical detection of the $1_0 - 0_0$ ground-state transition of ammonia, \nhhh, which near 572 GHz is also unobservable from the ground due to absorption from the 557 GHz \hzo\ line's broad line wings.

Here we report the detection of the \hzo\ and the \nhhh\ ortho-ground-state rotational lines toward PKS 1830$-$211 with the APEX telescope\footnote{APEX observations of the \hzo\ line toward PKS 1830$-$211 have been announced previously by \citet{WiklindCombes2005}}. In \S\ref{obs} we describe the observations and data reduction. Our results are presented in \S\ref{res} and their analysis in \S\ref{analysis}. Astrochemical implications of our study are discussed in \ref{abh2o}.

\section{\label{obs}Observations and data reduction}
Our observations  were made in 2007 July 2--5 under generally good weather conditions
with APEX, the 12-m Atacama Pathfinder Experiment telescope \citep{Gusten_etal2006}.
Opacities at the observing frequencies ($\approx 300$ GHz) were between 0.02 and 0.12 throughout the observations and (single sideband) system temperature between 240 and 390 K.
We observed the ortho-ground state lines of water (\hzo) and  ammonia (\nhhh) ,
namely the \hzo\ $J_{K_{a}K_{c}} = 1_{10} - 1_{01}$ and the \nhhh\
$J_K = 1_0 - 0_0$
transitions. These were redshifted from their rest frequencies of $556936.00\pm0.05$ and $572498.1\pm0.3$ MHz to observing
frequencies, $\nu$, of 295328.3 and 303580.4 and MHz, respectively.
We assumed the redshift accurately determined from mm-/cm-wavelength
absorption measurements, namely $z = 0.88582$. All spectra are presented with
velocities in the heliocentric frame; add 12.1 \kms\ for LSR velocities.

Both lines were measured in the lower sideband (LSB) with
the APEX 2a facility receiver \citep{Risacher_etal2006}. Calibration was
obtained using the chopper wheel technique, considering the
different atmospheric opacities in the signal and image sidebands of the
employed double sideband receivers. To ensure
flat spectral baselines, the newly available wobbling secondary was chopped with a frequency of
1.5 Hz and a throw of $100''$ about the cross elevation axis. The wobbler was operated in symmetric mode, which means that source and off position are interchanged between subsequent subscans, which cancels any asymmetries in the optical paths.
Such observations deliver a reliable estimate of the continuum level.
The radiation was analyzed with the MPIfR Fast Fourier Transform spectrometer,
which provides 16384 frequency channels  over the 1\,GHz intermediate frequency
bandwidth \citep{Klein_etal2006}. To increase the signal to noise ratio, the
spectra were smoothed to effective velocity resolutions appropriate for the
measured linewidths, i.e., 4.0 and 3.9 \kms\ for the \hzo\ and \nhhh\ lines, respectively. To check the telescope pointing,  drift scans were made across the strong continuum sources Sgr B2 N, G10.47+0.03, Jupiter, and  Mars.
Pointing
corrections were derived from these measurements. The pointing
was found to be accurate to within $\approx 3''$, acceptable
given the FWHM beam size, $\theta_{\rm B}$, which is $21''$ FWHM at 300 GHz.

\begin{figure}[t]
\begin{center}
\includegraphics[width=7.5cm,angle=0]{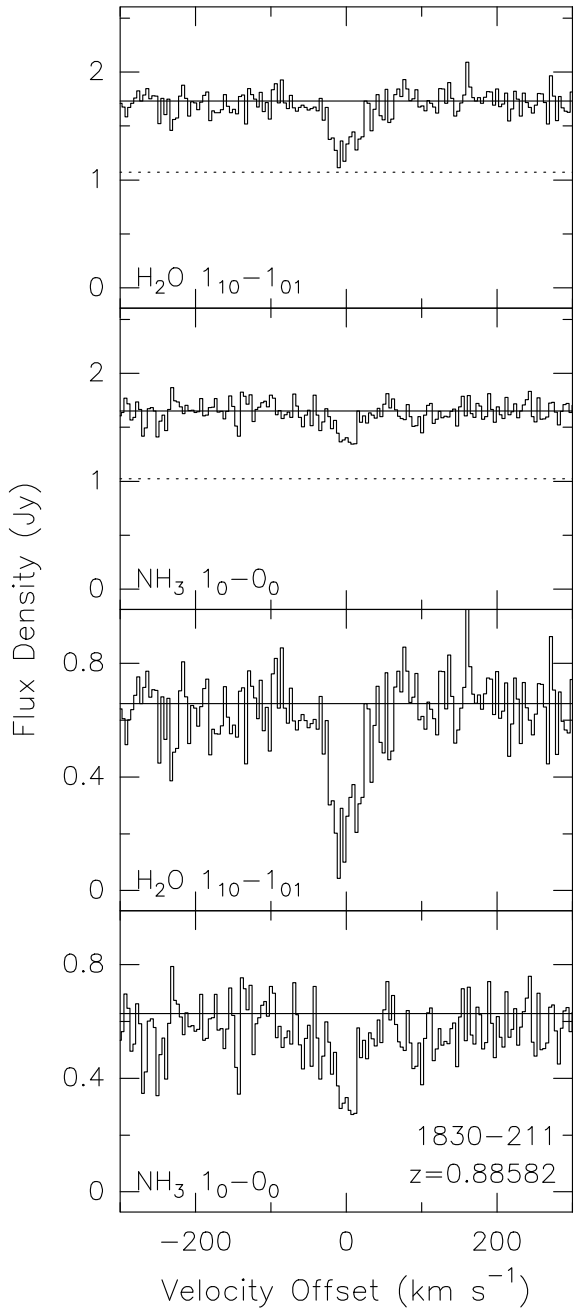}
\caption{\label{H2OandNH3Spectra} Spectra of absorption
in the $J_{K_{a}K_{c}} = 1_{10} - 1_{01}$\ transition
of water vapor  and the $J_K = 1_0 - 0_0$\ transition
of ammonia  taken with the APEX telescope
toward PKS 1830$-$211. The velocity scale is in the
heliocentric frame and assumes a redshift of 0.88582.
For the two upper spectra, the continuous horizontal line gives
the continuum level discussed in \S\ref{obs} while the
difference between the dotted and continuous lines gives the actually absorbed continuum level from the SW component (total covering factor 0.3; see \S\ref{ods}). For clarity, in the lower panels the H$_2$O and \nhhh\ spectra are shown relative to the actually absorbed continuum level.}
\end{center}
\end{figure}

We have converted our line intensities, measured with the chopper wheel technique,
into a flux density, $S$, scale (i.e., in Jansky units), assuming the 
aperture efficiency observationally determined by \citet{Gusten_etal2006}
at 352 GHz and higher frequencies extrapolated to 300 GHz; 1 K of antenna temperature corrected for above the atmosphere corresponds to 37.7 Jy.
We present our observed spectra in Fig. \ref{H2OandNH3Spectra} with the continuum levels indicated (see \ref{cont}). No baselines were subtracted from these spectra.

\section{\label{res}Results}
\subsection{\label{cont}Continuum intensities}
In the double-sideband APEX 2a receiver, the two sidebands are 12 GHz apart
and continuum emission from both sidebands is detected. The effective center
frequencies for the determination of the continuum levels are 301.3 and 309.6 GHz, respectively, for the two sidebands.
We determine continuum levels, $S_{\rm C}$, of 1.73 Jy at the
former and 1.65 Jy at
the latter frequency, respectively. Given the absolute calibration uncertainties,
which we estimate at 10\%, the difference between the two measurements
is 
not
significant.
Keeping in mind that PKS~1830$-$211 is highly variable, we note that our
flux densities are higher than the 1.3 Jy measured with the
Large APEX Bolometer Camera (LABOCA) shortly before our observations, i.e.,
in 2007 May (A. Weiss and F. Schuller, personal communication).
The LABOCA's  effective (filter-averaged) frequency is 345 GHz, 49 GHz higher
than the H$_2$O line frequency. If PKS 1830$-$211's submillimeter flux
did \textit{not}
vary between then and the time of our measurements, comparison of our and the LABOCA
data shows that
the trend
of dropping flux with increasing frequency continues at least up to 345 GHz.

We estimate the uncertainties of our absolute flux density
calibration and the systematic uncertainties in the optical depth
determinations (see \S\ref{ods}) to be $\approx 10$\%. This means
that the uncertainties in the optical depth determinations are dominated
by the limited signal-to-noise ratio  of the line measurements.

\subsection{\label{mrcf}The magnification ratio and covering factor}
With our $\approx 20''$ FWHM beam we are detecting the sum of the
continuum flux densities from the NE and the SW images. However, from
interferometric imaging it is clear that observed absorption at zero velocity
(relative to $z=0.88582$) only arises toward the weaker SW image
\citep{Frye1997, Menten1999, Muller2006}. Thus, we must correct the
observed apparent optical depths to true optical depths
taking this into account. We here define the ``magnification ratio'',
$\mathfrak{R}$, for PKS 1830$-$211 as the ratio of the radio or (sub)millimeter
flux densities of the stronger NE to the weaker SW image. Due to time delay
between the two ray paths, this ratio is temporally variable, but if one
accounts for that, $\mathfrak{R} \approx 1.5$ has been estimated in the radio
regime \citep{Lovell1998}. Observations with the IRAM Plateau de Bure Interferometr (PdBI) taken 
between 1995 and 2003 in the 3 mm wavelength range
show $\mathfrak{R} \approx 1.5$ -- 1.75 \citep{Muller2006}, with a best
value of 1.66. 
The flux density of the SW image is $\eta_{\rm cov}S_{\rm C} \equiv S_{\rm C}/(1 + \mathfrak{R})$. Note that here
the ``covering factor'' $\eta_{\rm cov}$, which accounts for the
fact that only the SW image is absorbed, ranges from 0.36 to 0.40, corresponding to the range in $\mathfrak{R}$ given above.
What fraction of that image's
few mas/few pc size area is absorbed is not considered and is assumed
to be near unity as suggested by Very Long Baseline Array (VLBA)
observations \citep{Carilli1998}.

\subsection{\label{ods}Observed and corrected optical depths}
A Gaussian fit to the \hzo\ line yields a peak line signal,
$S_{\rm L}$, of  
$-0.47 (0.11)$ Jy, and an integrated line signal, $\int S_{\rm L}$dv, of $-22.1 (2.0)$~Jy~\kms,
an FWHM linewidth, $\Delta$v, of 44 (5) \kms\ and
velocity offset, v, of $-1.8 (2.0)$ \kms\ relative to $z = 0.88582$.
Using this line and the continuum flux density derived in \S\ref{cont} (10\%\ uncertainty for the latter),  we calculate an apparent optical depth,
\begin{equation}
\tau_{\rm app} = -ln(1 - |S_{\rm L}|/S_{\rm C})
\end{equation}
of $0.32^{+15}_{-11}$ and a velocity-integrated
apparent optical depth, $\smallint \tau_{\rm app}~d{\rm v}$ of 15.0 km~s$^{-1}$.

For the \nhhh\ line we find $S_{\rm L} = -0.29 (6)$ Jy, $\int S_{\rm L}$dv =  $-10.4 (1.6)$~Jy~\kms,
$\Delta$v = 34 (7) \kms, and v $= 0.5 (2.5)$ \kms, yielding $\tau_{\rm app} = 0.19^{+0.08}_{-0.06}$
and $\smallint \tau_{\rm app}~d{\rm v} = 6.4$ km~s$^{-1}$.

Replacing $S_{\rm C}$ with $\eta_{\rm cov} S_{\rm C}$ in the above equation and considering that $0.36 < \eta_{\rm cov} <0.40)$, we obtain
a true optical depth of$1.3^{+\infty}_{-0.6}$ and 61 km~s$^{-1}$ for its velocity-integrated value. This means that from our data we cannot give an upper bound for the line's opacity.

For the \nhhh\ line we calculate 
$0.62^{+0.46}_{-0.24}$
and 22 km~s$^{-1}$ for the true optical depth and its velocity-integrated value, respectively.

For a spectral line from an upper level $u$, with energy $E_u$,  to a lower level $l$, with
energy $E_l$, the column density in the upper level is
\begin{equation}
N_u = {{8\pi\nu^3}\over{c^3A_{ul}}}\big[e^{h\nu/kT_{\rm ex}} - 1\big]^{-1}\smallint \tau d{\rm v},
\end{equation}
\noindent
where $T_{\rm ex}$ is the excitation temperature, $A_{ul}$ the Einstein A coefficient;
$h$ and $k$ are the Planck and Boltzmann constants,
respectively, and $c$ is the speed of light. $\nu = (E_u - E_l)/h$ is the line's rest frequency. In convenient units,
\begin{equation}\label{nupper}
N_u ({\rm cm}^{-2})= 93.28~{{\nu^3({\rm GHz})}\over{A_{ul}({\rm s}^{-1})}}
\big[e^{0.048~\nu({\rm GHz})/T_{\rm ex}({\rm K})}-1\big]^{-1}\smallint\tau d{\rm v~(km~s}^{-1})
\end{equation}

The total column density is given by
\begin{equation}\label{ntot}
N_{\rm rot}= {{N_u}\over{g_u}}e^{E_u/kT_{\rm rot}}Q(T_{\rm rot}),
\end{equation}
where $g_u$ is the line's upper level degeneracy and $Q$ is the partition function
for the rotation temperature, $T_{\rm rot}$. We assume
Local Thermodynamic Equilibrium (LTE), which means $T_{\rm rot} = T_{\rm ex}$.

\section{\label{analysis}Analysis}
\subsection{\label{h2oabs}Water vapor absorption}
For the \hzo\ $J_{K_{a}K_{c}} = 1_{10} - 1_{01}$ transition, $g_u = g_l = 9$
and $A_{ul} = 3.45~10^{-3}$~s$^{-1}$, where $u$ and $l$ denote the $1_{10}$
and $1_{01}$ energy levels, respectively. To consider the state of
thermalization of this line, we examine the rates for collisional
deexcitation from the $1_{01}$ level, which are accessible at the website
of the Leiden Atomic and Molecular Database (LAMDA)\footnote{http://www.strw.leidenuniv.nl/~moldata/} \citep{Schoier2005}.
The calculations of \citet{Grosjean2003} for collisions
between ortho-H$_2$ and ortho-\hzo\ at low temperatures (5--20 K),
extended to higher temperatures (140 K) by \citet{Phillips1996},
yield rate coefficients for transitions from the upper $1_{10}$
to the lower $1_{01}$ level, $\gamma_{ul}$, between $8.2~10^{-12}$ and $2.2~10^{-11}$~cm$^3$s$^{-1}$ for kinetic temperatures between 5 and 80 K, where the latter number applies for the bulk of the gas from which absorption in the ammonia inversion transitions is also observed \citep{Henkel2008}.
This implies a critical density, $n_{\rm crit} = A_{ul}/\gamma_{ul}$, of several times $10^{8}$~cm$^{3}$, which is very much higher than any plausible value of the hydrogen density of the absorbing material. It is therefore straightforward to assume that the \hzo\ level populations are thermalized at the temperature of the cosmic microwave background, $T_{\rm CMB}$.
In other words, $T_{\rm ex}$ in Eqn. \ref{nupper}  is equal to  $T_{\rm CMB} = (1+z)~2.728$~ K = 5.145 K.
Using these values and the lower limit to integrated optical depth 
discussed in \S\ref{ods}, i.e. 0.7 km~s${-1}$, with $T_{\rm ex} = T_{\rm CMB}$,  Eqn. \ref{nupper} gives
an upper level column density of $N_{1_{10}} = 5.6~10^{11}$~cm$^{-2}$.

Noting that the ortho-\hzo\ partition function for this
temperature, $Q_{{\rm o-H}_2{\rm O}}(5.145 K)$, is $\approx 9.05$,
we can use Eqn. \ref{ntot} to find that the total ortho-\hzo\
column density, $N($o-H$_2$O), is  $9.9~10^{13}$~cm$^{-2}$, 182.14 times larger than
$N_{10}$ and, from detailed balance, that
almost all ortho-\hzo\ molecules (99.45\%) are in the $1_{01}$ ground-state level!
To verify this, we used  the non-LTE radiative transfer program
RADEX\footnote{http://www.strw.leidenuniv.nl/~moldata/radex.html}
created and made available to the community by \citet{VanderTak2007}
to perform statistical equilibrium calculations. For a set of input
excitation conditions, namely background temperature (5.145 K),
kinetic temperature, and H$_2$ density, RADEX calculates level
populations with column density and line width as radiative transfer
input parameters and returns the excitation temperatures and optical
depths of user-selected spectral lines.

Running RADEX, we find that for densities up to $10^6$ \ccm\ the
excitation temperature does not become significantly larger than
$T_{\rm CMB}$. For the line width derived in \S\ref{res}
RADEX returns the optical depth value discussed above ($\tau = 0.7$)
for an ortho-\hzo\ column density of $9.3~10^{13}$ cm$^{-2}$,
close to above estimate.
This result is independent of all plausible values of the
kinetic temperature and densities up to $10^6$~cm$^{-3}$.

We emphasize that the above estimate is calculated for the lower limit of the true integrated optical depth derived
in \S\ref{ods} 
and a strict lower limit. The 
upper uncertainty bound of the optical depth is not determinable from our noisy data.


\subsection{\label{nh3abs}Ammonia absorption}
For the \nhhh\ $J_K = 1_0 - 0_0$\ transition, $g_u = 12$\ and for the  $(J,K) = (1,1)$
transition $g_u = 6$. These values include factors for nuclear spin degeneracies.
The very different Einstein A coefficients for these lines are $A_{1_0-0_0} = 1.61~10^{-3}$~s$^{-1}$ and $A_{1,1} = 1.71~10^{-7}$~s$^{-1}$, respectively.

Deexcitation rates for collisions of H$_2$ with ortho-NH$_3$
have been calculated by \citet{Danby_etal1988} and are also
accessible at the LAMDA website. We find that the rate for
collisional deexcitation, $\gamma_{01}$, from the $1_0$ to the $0_0$ level for temperatures between 15 and 300 K varies little and is between $4.3~10^{-11}$ and $6.2~10^{-11}$~cm$^3$s$^{-1}$.
The critical density of the $J_K = 1_0 - 0_0$\ transition, $n_{\rm crit}(1_0-0_0) = A_{ul}/\gamma_{ul}$, is, thus, $\sim 3~10^7$~cm$^{-3}$, which is, as in the case of the \hzo\ line discussed above, many orders of magnitude higher than the densities (of order a few times $10^3$~cm$^{-1}$)  of the absorbing interstellar medium (ISM) in PKS 1830$-$211. It is therefore also clear that the population of the \nhhh\ $1_0$ relative to the $0_0$ level is controlled by the cosmic microwave background temperature. Using Eqn. \ref{nupper}, we calculate $N_{1_0} = 1.4~10^{12}$~cm$^{-2}$ for the column density of the $1_0$ level.
We can calculate the ortho-\nhhh\ partition function by direct summation with parameters from LAMDA and get, assuming $T_{\rm rot} = 5.145$ K, $Q_{{\rm o-NH}_3}(5.145 {\rm K})\approx 4.06$. Using Eqn. \ref{ntot} we find that the total ortho-\nhhh\ column density, $N($o-NH$_3)$, is $9.1~10^{13}$~cm$^{-2}$, 70 times larger than $N_{10}$.

Again using RADEX, we find that to reproduce our observed opacity (0.62 at a line width
of 34 \kms) requires an ortho-ammonia column density, $N$(o-NH$_3)$, of $1.3~10^{14}$ cm$^{-2}$.
For $T_{\rm rot}=5.145$ K, a quarter of all  ortho-\nhhh\ molecules  are in the $0_{0}$ ground-state level while, in contrast to o-\hzo, the majority of molecules are in other energy levels.
We used a kinetic temperature of 80 K \citep[see \S\ref{intro} and ][]{Henkel2008} but note that lower temperatures would only imply a 20\%\ lower column density at the most.

\section{\label{abh2o}Discussion - Abundances and astrochemical implications}
To calculate molecular abundances relative to molecular hydrogen, we
assume an H$_2$ column density of $3~10^{22}$~cm$^{-1}$ for the absorbing material toward PKS 1830$-$211-SW. This value was estimated by \citet{WiklindCombes1996} assuming a CO to H$_2$ ratio of $10^{-4}$.
It is somewhat larger than the  $(1.8\pm0.3)~10^{22}$~cm$^{-2}$ derived from
X-ray observations with the ROSAT satellite \citep{MathurNair1997}.

With the  lower limit to $N($o-H$_2$O) determined above,  $9.3~10^{13}$ cm$^{-2}$ determined above,
we find that the ortho-H$_2$O to
H$_2$ abundance ratio, $X$(o-H$_2$O), is greater than $\approx 4~10^{-9}$.
To put this value into context, SWAS observed the same
H$_2$O line in a wide variety of Galactic sources
\citep[for an overview see ][]{Melnick2000}. \citet{Snell2000a}
were able to determine  H$_2$O abundances, with values similar
to our PKS 1830$-$211 number, toward dense quiescent regions
of molecular clouds (between $6~10^{-10}$ and $1~10^{-8}$), with values
near the higher number found in the core regions of Giant Molecular Clouds
\citep[GMCs; see also][]{Bergin2000}.

Several observational effects have to be considered concerning the
SWAS results: The regions toward which SWAS detected the H$_2$O line at
low abundance are core regions of molecular clouds, some of them GMCs,
i.e., regions of enhanced density and thus,
enhanced column density, producing high enough optical depths to make
the line detectable. They are quiescent in the sense they don't harbor
outflows.
Whereas toward the spatially confined regions harboring outflows,
as evidenced by broad line wings and other tracers, SWAS finds
several orders of magnitude higher H$_2$O abundances \citep{Bergin2000, Bergin2003}. Similarly, higher H$_2$O abundances are found near the warmer interfaces to HII regions \citep{Snell2000b}. SWAS gives little information, just rough upper limits, on water in the lower density extended gas of giant molecular clouds, simply because the expected line intensities are  prohibitively low. Similarly, only coarse upper limits (around $10^{-7}$--$10^{-6}$ relative to H$_2$) can be derived toward cold dark clouds such as TMC-1, L134N, or B335, which also have lower  average densities ($~10^4$~cm$^{-4}$) than warm GMC cores. Moreover, their low temperatures ($~10~$K) are not conducive for producing emission in the $1_{10}-1_{01}$ line, whose upper energy level is 34 K above ground.

The size of the $\sim 4'$ FWHM beam of SWAS at the H$_2$O line
rest frequency (557 GHz) for a region at distance $D$ (measured in kpc)
corresponds to an area of size $\approx 1~{\rm pc} \times D({\rm kpc})$.
This means that for sources at a few kpc distance, similar size scales
are sampled as for PKS 1830$-$211 with the VLBA beam of \citet{Carilli1998},
which has a FWHM of 1 mas (7.5 pc at $z=0.89$). Indeed, the SWAS H$_2$O
spectrum of the W 33 region which, at a (near-kinematic) distance of 4 kpc,
has a width of $\sim 25$~km~s$^{-1}$ full width at zero power (FWZP)\footnote{Intriguingly, the W33 spectrum
shows absorption against the source's strong submillimeter continuum
emission in its central velocity range.}. This is very wide for
a quiescent cloud core which, as the low H$_2$O abundance suggests,
does not contain outflows. The width could be determined by a combination
of the presence of velocity gradients in the region unresolved by the
beam and the existence of emission from several spiral arms caused by the
line of sight crossing the plane of the Galaxy. Alternately, an outflow could be present, which actually is suggested by emission from the outflow tracer SiO from which lines of width comparable to the \hzo\ line's are observed toward W33 \citep{Miettinen2006}. In that case, the apparent low \hzo\ abundance could result from the superposition of a compact (i.e. beam diluted) higher-abundance outflow component and a more extended component of much lower abundance.

Even in comparison to the relatively broad \hzo\ emission toward W33, the 55--60~km~s$^{-1}$ FWZP width of the PKS 1830$-$211-SW absorption appears \textit{extraordinarily} broad, especially since we are
viewing the lensing galaxy close to face-on and the 2 milliarcsecond
size absorption ``pencil beam'' afforded by the size of the background source
corresponds to a scale of 15 pc, smaller than a single GMC.
Interferometric observations of the local face-on spiral M51
with a resolution corresponding to a much larger $\sim150$ pc
reveal linewidths only between 10 and 20 \kms\ in the galaxy's spiral
arms \citep{Aalto1999}.

The probability of a random line of sight passing through a cloud
core certainly is lower than passing through a lower density,
more ``normal'', region of a GMC. The \hzo\ abundance in the bulk
volume of molecular clouds is essentially unknown and, for the
above reasons, SWAS or ODIN cannot deliver very meaningful limits.
As the multi-molecule studies of galactic absorption against
extragalactic radio sources conducted by H. Liszt and R. Lucas
have persuasively demonstrated, 
all the other molecules
found toward PKS 1830$-$211 can be detected in general molecular ISM  gas
\citep{Lucas2000, Liszt2001, Liszt2004a, Liszt2004b, Liszt2006}.
However, as explained in the following, the gas absorbing PKS
1830$-$211 appears to have a much higher temperature.

Using the NRAO Green Bank Telescope, \citet{Henkel2008} detected absorption in the \nhhh\ $(J,K)$
inversion lines toward PKS 1830$-$211 for  $J = K = 1$ up to 10, which arise from levels
with energies between 23 and 1036 K above ground. Fig. \ref{BothNH3Spectra}
compares our spectrum of the $J_K = 1_1-0_0$ line with their $J,K = (1,1)$
spectrum. There appears to be excess absorption in the former at
velocities $> 0$ \kms, although the signal-to-noise ratio is rather
limited. If real, this would be difficult to explain since, given that
the absorbed region is larger at the redshifted (1,1) frequency (12.56 GHz)
than at the $1_1-0_0$ frequency, one would expect the opposite. Higher
signal to noise data obtainable with the Submillimeter Array seem highly
desirable!

\citet{Henkel2008} find that to explain the observed inversion line optical depths,
80--90\%\ of the NH$_3$ must be in warm (80 K) gas with a total ortho-
plus para-NH$_3$ column density of $5~10^{14}$~cm$^{-2}$, if the excitation
temperature (assumed to be the same for all lines) equals $T_{\rm CMB}$.
When modeling the clouds in their SWAS sample, \citet{Snell2000b} assumed
temperatures between 25 and 40 K for the GMC cores, as determined from
observations of other tracers, which is significantly lower than the 80 K
quoted above.
Moreover, the kinetic temperatures of 25--30 K derived by \citet{Liszt2006}
from the \nhhh\ (1,1) and (2,2) lines are substantially lower. This poses an
interesting question: Is the $z=0.89$ absorbing cloud much hotter than
Galactic clouds or could a dilute warm molecular component in the Milky Way
(and other galaxies) so far have escaped observational attention?

The ortho-\nhhh\ column density determined by us is a factor of 2 lower
than the value derived from inversion lines \citep{Henkel2008}. Interestingly, a \textit{much larger} discrepancy
seems to exist between the \nhhh\ column density derived from the
Odin $1_1-0_0$ spectrum \citep{Liseau2003} and the much higher value
implied by the (1,1) and (2,2) inversion lines mapped by \citet{Zeng1984} toward the dark cloud $\rho$~Oph A.

\section{Outlook -- constraints on variation of fundamental constants}
Comparing the redshifts of various spectral lines measured toward the
same high-$z$ object can in principle yield constraints on possible
temporal variations of fundamental constants over significant parts
of the age of the Universe \citep[see, e.g. ][]{Curran2004}. A precondition
is that the lines in question have different dependencies on the constants,
e.g., the fine structure constant, $\alpha$, and/or the ratio of the mass
of the proton to that of the electron, $\mu = m_{\rm p}/m_{\rm e}$.
Traditionally exploited with optical/UV quasar absorption lines, the
method has recently been applied to radio and millimeter absorption
lines toward, amongst others, the gravitational lens systems PMN J0134$-$0931
\citep{Kanekar2005} and B0218+357 \citep{Murphy2008}.

The ammonia molecule is especially interesting with respect
to this, as its inversion transitions are strongly dependent on $\mu$:
\citet{FlambaumKozlov2007} find that the fractional change in frequency
of these transitions
${\delta \nu_{\rm inv}}/{\nu_{\rm inv}} = {\delta z}/(1+z)$
is given by $-4.46{\delta \mu}/{\mu}$ or,
in other words,  ${\Delta \nu_{\rm inv}} \propto \Delta \mu^{-4.46}$.
For fractional variations of rotational line frequencies, $\nu_{\rm rot}$ ,
in contrast, the proportionality factor is unity,
i.e., $\delta \nu_{\rm rot}/\nu_{\rm rot} = -{\delta \mu}/{\mu}$.
Comparing \nhhh\ inversion line redshifts with redshifts of rotational
lines should thus provide sensitive limits on the variation of $\mu$.

\citet{Murphy2008} compare the measured redshifts of the
\nhhh\ $(J, K) = $ $(1,1), (2,2)$ and $(3,3)$ lines observed
toward B0218+357 by \citet{Henkel2005} with rotational lines
from HCO$^{+}$ and HCN  and obtain $\delta \mu/{\mu} = (0.74 \pm 0.47)~10^{-6}$
over a look-back time of 6.2 Gyr, corresponding to the absorber's
redshift of 0.68466. One fundamental limitation of this measurement
is the uncertainty over the two species' possibly different distributions
leading to different covering factors. Different covering factors might
also result from source size variations between 14 GHz (\nhhh) and 105 GHz
(HCO$^+$ and HCN).
In addition, lines from different species might also cover different
velocity ranges.

The ortho-ground-state rotational transition of \nhhh\ we have detected
toward PKS 1830$-$211 can in principle deliver a more accurate and meaningful
comparison between the redshifts of a rotational line redshift
and of the inversion lines, which have been determined by
\cite{Henkel2008}; see \S\ref{intro}).
The velocity accuracy of our APEX measurement
(2.5 \kms) translates into a redshift uncertainty,
$\delta z$, of $1.6~10^{-5}$  and a value of
${\delta \mu}/{\mu} \approx 0.224~{\delta z}/(1+z)$ of $1.9~10^{-6}$,
which is three times larger than the value derived by \citet{Murphy2008}
for B0218+357.

\begin{figure}[t]
\begin{center}
\includegraphics[width=7.5cm,angle=0]{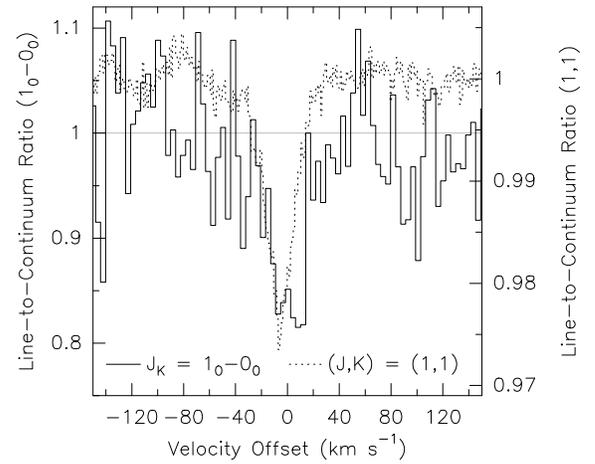}
\caption{\label{BothNH3Spectra} Ammonia absorption spectra of the  $J_K = 1_0 - 0_0$\
rotational transition (full line) and the $(J,K) = (1,1)$
inversion transition (dashed line) toward PKS 1830$-$211.
The velocity scale is in the heliocentric frame and assumes
a redshift of 0.88582. The y-axes give the line-to-continuum
ratio for the $1_0 - 0_0$ line  (left), for which the continuum
level is 1.65 Jy, and the (1,1) line (right) for which the continuum level is 8.2 Jy.}
\end{center}
\end{figure}

Our \nhhh\ $1_0-0_0$ spectrum is of very limited quality and even looks
quite different from the (1,1) spectrum (see Fig. \ref{BothNH3Spectra}),
although that can be attributed to the former line's poor signal-to-noise
ratio. A few hour long observation with the Submillimeter Array
\citep{Ho2004} will result in a much improved spectrum, which,
together with the (1,1) spectrum 
could lead to a 
significant advancement in
investigating the  variation of $\mu$ over the past 7.2 Gyr.

\acknowledgements{We would like to thank Gary Melnick for useful comments on this paper.}

\bibliographystyle{aa}
\bibliography{1830}

\begin{thebibliography}{56}
\expandafter\ifx\csname natexlab\endcsname\relax\def\natexlab#1{#1}\fi

\bibitem[{{Aalto} {et~al.}(1999){Aalto}, {H{\"u}ttemeister}, {Scoville}, \&
  {Thaddeus}}]{Aalto1999}
{Aalto}, S., {H{\"u}ttemeister}, S., {Scoville}, N.~Z., \& {Thaddeus}, P. 1999,
  \apj, 522, 165

\bibitem[{{Bergin} {et~al.}(2003){Bergin}, {Kaufman}, {Melnick}, {Snell}, \&
  {Howe}}]{Bergin2003}
{Bergin}, E.~A., {Kaufman}, M.~J., {Melnick}, G.~J., {Snell}, R.~L., \& {Howe},
  J.~E. 2003, \apj, 582, 830

\bibitem[{{Bergin} {et~al.}(2000){Bergin}, {Melnick}, {Stauffer}, {Ashby},
  {Chin}, {Erickson}, {Goldsmith}, {Harwit}, {Howe}, {Kleiner}, {Koch},
  {Neufeld}, {Patten}, {Plume}, {Schieder}, {Snell}, {Tolls}, {Wang},
  {Winnewisser}, \& {Zhang}}]{Bergin2000}
{Bergin}, E.~A., {Melnick}, G.~J., {Stauffer}, J.~R., {et~al.} 2000, \apjl,
  539, L129

\bibitem[{{Bronfman} {et~al.}(1988){Bronfman}, {Cohen}, {Alvarez}, {May}, \&
  {Thaddeus}}]{Bronfman1988}
{Bronfman}, L., {Cohen}, R.~S., {Alvarez}, H., {May}, J., \& {Thaddeus}, P.
  1988, \apj, 324, 248

\bibitem[{{Carilli} {et~al.}(1998){Carilli}, {Menten}, {Reid}, {Rupen}, \&
  {Claussen}}]{Carilli1998}
{Carilli}, C.~L., {Menten}, K.~M., {Reid}, M.~J., {Rupen}, M., \& {Claussen},
  M. 1998, in Astronomical Society of the Pacific Conference Series, Vol. 144,
  IAU Colloq. 164: Radio Emission from Galactic and Extragalactic Compact
  Sources, ed. J.~A. {Zensus}, G.~B. {Taylor}, \& J.~M. {Wrobel}, 317

\bibitem[{{Chengalur} {et~al.}(1999){Chengalur}, {de Bruyn}, \&
  {Narasimha}}]{Chengalur1999}
{Chengalur}, J.~N., {de Bruyn}, A.~G., \& {Narasimha}, D. 1999, \aap, 343, L79

\bibitem[{{Combes} \& {Wiklind}(1997)}]{CombesWiklind1997}
{Combes}, F. \& {Wiklind}, T. 1997, \apjl, 486, L79

\bibitem[{{Courbin} {et~al.}(2002){Courbin}, {Meylan}, {Kneib}, \&
  {Lidman}}]{Courbin2002}
{Courbin}, F., {Meylan}, G., {Kneib}, J.-P., \& {Lidman}, C. 2002, \apj, 575,
  95

\bibitem[{{Curran} {et~al.}(2004){Curran}, {Kanekar}, \&
  {Darling}}]{Curran2004}
{Curran}, S.~J., {Kanekar}, N., \& {Darling}, J.~K. 2004, New Astronomy Review,
  48, 1095

\bibitem[{{Danby} {et~al.}(1988){Danby}, {Flower}, {Valiron}, {Schilke}, \&
  {Walmsley}}]{Danby_etal1988}
{Danby}, G., {Flower}, D.~R., {Valiron}, P., {Schilke}, P., \& {Walmsley},
  C.~M. 1988, \mnras, 235, 229

\bibitem[{{Flambaum} \& {Kozlov}(2007)}]{FlambaumKozlov2007}
{Flambaum}, V.~V. \& {Kozlov}, M.~G. 2007, Physical Review Letters, 98, 240801

\bibitem[{{Frye} {et~al.}(1997){Frye}, {Welch}, \& {Broadhurst}}]{Frye1997}
{Frye}, B., {Welch}, W.~J., \& {Broadhurst}, T. 1997, \apjl, 478, L25

\bibitem[{{Grosjean} {et~al.}(2003){Grosjean}, {Dubernet}, \&
  {Ceccarelli}}]{Grosjean2003}
{Grosjean}, A., {Dubernet}, M.-L., \& {Ceccarelli}, C. 2003, \aap, 408, 1197

\bibitem[{{G{\"u}sten} {et~al.}(2006){G{\"u}sten}, {Nyman}, {Schilke},
  {Menten}, {Cesarsky}, \& {Booth}}]{Gusten_etal2006}
{G{\"u}sten}, R., {Nyman}, L.~{\AA}., {Schilke}, P., {et~al.} 2006, \aap, 454,
  L13

\bibitem[{{Henkel} {et~al.}(2008){Henkel}, {Braatz}, {Menten}, \&
  {Ott}}]{Henkel2008}
{Henkel}, C., {Braatz}, J.~A., {Menten}, K.~M., \& {Ott}, J. 2008, \aap, 485,
  451

\bibitem[{{Henkel} {et~al.}(2005){Henkel}, {Jethava}, {Kraus}, {Menten},
  {Carilli}, {Grasshoff}, {Lubowich}, \& {Reid}}]{Henkel2005}
{Henkel}, C., {Jethava}, N., {Kraus}, A., {et~al.} 2005, \aap, 440, 893

\bibitem[{{Hjalmarson} {et~al.}(2003){Hjalmarson}, {Frisk}, {Olberg},
  {Bergman}, {Bernath}, {Biver}, {Black}, {Booth}, {Buat}, {Crovisier},
  {Curry}, {Dahlgren}, {Encrenaz}, {Falgarone}, {Feldman}, {Fich},
  {Flor{\'e}n}, {Fredrixon}, {Gerin}, {Gregersen}, {Hagstr{\"o}m}, {Harju},
  {Hasegawa}, {Horellou}, {Johansson}, {Kyr{\"o}l{\"a}}, {Kwok}, {Larsson},
  {Lecacheux}, {Liljestr{\"o}m}, {Lindqvist}, {Liseau}, {Llewellyn}, {Mattila},
  {M{\'e}gie}, {Mitchell}, {Murtagh}, {Nyman}, {Nordh}, {Olofsson}, {Olofsson},
  {Olofsson}, {Pagani}, {Persson}, {Plume}, {Rickman}, {Ristorcelli},
  {Rydbeck}, {Sandqvist}, {von Sch{\'e}ele}, {Serra}, {Torchinsky}, {Tothill},
  {Volk}, {Wiklind}, {Wilson}, {Winnberg}, \& {Witt}}]{Hjalmarson2003}
{Hjalmarson}, {\AA}., {Frisk}, U., {Olberg}, M., {et~al.} 2003, \aap, 402, L39

\bibitem[{{Ho} {et~al.}(2004){Ho}, {Moran}, \& {Lo}}]{Ho2004}
{Ho}, P.~T.~P., {Moran}, J.~M., \& {Lo}, K.~Y. 2004, \apjl, 616, L1

\bibitem[{{Huettemeister} {et~al.}(1995){Huettemeister}, {Wilson},
  {Mauersberger}, {Lemme}, {Dahmen}, \& {Henkel}}]{Huettemeister1995}
{Huettemeister}, S., {Wilson}, T.~L., {Mauersberger}, R., {et~al.} 1995, \aap,
  294, 667

\bibitem[{{Jauncey} {et~al.}(1991){Jauncey}, {Reynolds}, {Tzioumis}, {Muxlow},
  {Perley}, {Murphy}, {Preston}, {King}, {Patnaik}, {Jones}, {Meier}, {Bird},
  {Blair}, {Bunton}, {Clay}, {Costa}, {Duncan}, {Ferris}, {Gough}, {Hamilton},
  {Hoard}, {Kemball}, {Kesteven}, {Lobdell}, {Luiten}, {Mcculloch}, {Murray},
  {Nicholson}, {Rao}, {Savage}, {Sinclair}, {Skjerve}, {Taaffe}, {Wark}, \&
  {White}}]{Jauncey1991}
{Jauncey}, D.~L., {Reynolds}, J.~E., {Tzioumis}, A.~K., {et~al.} 1991, \nat,
  352, 132

\bibitem[{{Jin} {et~al.}(2003){Jin}, {Garrett}, {Nair}, {Porcas}, {Patnaik}, \&
  {Nan}}]{Jin2003}
{Jin}, C., {Garrett}, M.~A., {Nair}, S., {et~al.} 2003, \mnras, 340, 1309

\bibitem[{{Kanekar} {et~al.}(2005){Kanekar}, {Carilli}, {Langston}, {Rocha},
  {Combes}, {Subrahmanyan}, {Stocke}, {Menten}, {Briggs}, \&
  {Wiklind}}]{Kanekar2005}
{Kanekar}, N., {Carilli}, C.~L., {Langston}, G.~I., {et~al.} 2005, Physical
  Review Letters, 95, 261301

\bibitem[{{Klein} {et~al.}(2006){Klein}, {Philipp}, {Kr{\"a}mer}, {Kasemann},
  {G{\"u}sten}, \& {Menten}}]{Klein_etal2006}
{Klein}, B., {Philipp}, S.~D., {Kr{\"a}mer}, I., {et~al.} 2006, \aap, 454, L29

\bibitem[{{Koopmans} \& {de Bruyn}(2005)}]{KoopmansdeBruyn2005}
{Koopmans}, L.~V.~E. \& {de Bruyn}, A.~G. 2005, \mnras, 360, L6

\bibitem[{{Lidman} {et~al.}(1999){Lidman}, {Courbin}, {Meylan}, {Broadhurst},
  {Frye}, \& {Welch}}]{Lidman1999}
{Lidman}, C., {Courbin}, F., {Meylan}, G., {et~al.} 1999, \apjl, 514, L57

\bibitem[{{Liseau} {et~al.}(2003){Liseau}, {Larsson}, {Brandeker}, {Bergman},
  {Bernath}, {Black}, {Booth}, {Buat}, {Curry}, {Encrenaz}, {Falgarone},
  {Feldman}, {Fich}, {Flor{\'e}n}, {Frisk}, {Gerin}, {Gregersen}, {Harju},
  {Hasegawa}, {Hjalmarson}, {Johansson}, {Kwok}, {Lecacheux}, {Liljestr{\"o}m},
  {Mattila}, {Mitchell}, {Nordh}, {Olberg}, {Olofsson}, {Pagani}, {Plume},
  {Ristorcelli}, {Sandqvist}, {Sch{\'e}ele}, {Serra}, {Tothill}, {Volk}, \&
  {Wilson}}]{Liseau2003}
{Liseau}, R., {Larsson}, B., {Brandeker}, A., {et~al.} 2003, \aap, 402, L73

\bibitem[{{Liszt} \& {Lucas}(2001)}]{Liszt2001}
{Liszt}, H. \& {Lucas}, R. 2001, \aap, 370, 576

\bibitem[{{Liszt} \& {Lucas}(2004)}]{Liszt2004b}
{Liszt}, H. \& {Lucas}, R. 2004, \aap, 428, 445

\bibitem[{{Liszt} {et~al.}(2004){Liszt}, {Lucas}, \& {Black}}]{Liszt2004a}
{Liszt}, H., {Lucas}, R., \& {Black}, J.~H. 2004, \aap, 428, 117

\bibitem[{{Liszt} {et~al.}(2006){Liszt}, {Lucas}, \& {Pety}}]{Liszt2006}
{Liszt}, H.~S., {Lucas}, R., \& {Pety}, J. 2006, \aap, 448, 253

\bibitem[{{Lovell} {et~al.}(1998){Lovell}, {Jauncey}, {Reynolds}, {Wieringa},
  {King}, {Tzioumis}, {McCulloch}, \& {Edwards}}]{Lovell1998}
{Lovell}, J.~E.~J., {Jauncey}, D.~L., {Reynolds}, J.~E., {et~al.} 1998, \apjl,
  508, L51

\bibitem[{{Lovell} {et~al.}(1996){Lovell}, {Reynolds}, {Jauncey}, {Backus},
  {McCulloch}, {Sinclair}, {Wilson}, {Tzioumis}, {King}, {Gough}, {Ellingsen},
  {Phillips}, {Preston}, \& {Jones}}]{Lovell1996}
{Lovell}, J.~E.~J., {Reynolds}, J.~E., {Jauncey}, D.~L., {et~al.} 1996, \apjl,
  472, L5

\bibitem[{{Lucas} \& {Liszt}(2000)}]{Lucas2000}
{Lucas}, R. \& {Liszt}, H.~S. 2000, \aap, 358, 1069

\bibitem[{{Mathur} \& {Nair}(1997)}]{MathurNair1997}
{Mathur}, S. \& {Nair}, S. 1997, \apj, 484, 140

\bibitem[{{Melnick} {et~al.}(2000){Melnick}, {Stauffer}, {Ashby}, {Bergin},
  {Chin}, {Erickson}, {Goldsmith}, {Harwit}, {Howe}, {Kleiner}, {Koch},
  {Neufeld}, {Patten}, {Plume}, {Schieder}, {Snell}, {Tolls}, {Wang},
  {Winnewisser}, \& {Zhang}}]{Melnick2000}
{Melnick}, G.~J., {Stauffer}, J.~R., {Ashby}, M.~L.~N., {et~al.} 2000, \apjl,
  539, L77

\bibitem[{{Menten} {et~al.}(1999){Menten}, {Carilli}, \& {Reid}}]{Menten1999}
{Menten}, K.~M., {Carilli}, C.~L., \& {Reid}, M.~J. 1999, in Astronomical
  Society of the Pacific Conference Series, Vol. 156, Highly Redshifted Radio
  Lines, ed. C.~L. {Carilli}, S.~J.~E. {Radford}, K.~M. {Menten}, \& G.~I.
  {Langston}, 218

\bibitem[{{Meylan} {et~al.}(2005){Meylan}, {Courbin}, {Lidman}, {Kneib}, \&
  {Tacconi-Garman}}]{Meylan2005}
{Meylan}, G., {Courbin}, F., {Lidman}, C., {Kneib}, J.-P., \& {Tacconi-Garman},
  L.~E. 2005, \aap, 438, L37

\bibitem[{{Miettinen} {et~al.}(2006){Miettinen}, {Harju}, {Haikala}, \&
  {Pomr{\'e}n}}]{Miettinen2006}
{Miettinen}, O., {Harju}, J., {Haikala}, L.~K., \& {Pomr{\'e}n}, C. 2006, \aap,
  460, 721

\bibitem[{{Muller} {et~al.}(2006){Muller}, {Gu{\'e}lin}, {Dumke}, {Lucas}, \&
  {Combes}}]{Muller2006}
{Muller}, S., {Gu{\'e}lin}, M., {Dumke}, M., {Lucas}, R., \& {Combes}, F. 2006,
  \aap, 458, 417

\bibitem[{{Murphy} {et~al.}(2008){Murphy}, {Flambaum}, {Muller}, \&
  {Henkel}}]{Murphy2008}
{Murphy}, M.~T., {Flambaum}, V.~V., {Muller}, S., \& {Henkel}, C. 2008,
  Science, 320, 1611

\bibitem[{{Nordh} {et~al.}(2003){Nordh}, {von Sch{\'e}ele}, {Frisk}, {Ahola},
  {Booth}, {Encrenaz}, {Hjalmarson}, {Kendall}, {Kyr{\"o}l{\"a}}, {Kwok},
  {Lecacheux}, {Leppelmeier}, {Llewellyn}, {Mattila}, {M{\'e}gie}, {Murtagh},
  {Rougeron}, \& {Witt}}]{Nordh2003}
{Nordh}, H.~L., {von Sch{\'e}ele}, F., {Frisk}, U., {et~al.} 2003, \aap, 402,
  L21

\bibitem[{{Phillips} {et~al.}(1996){Phillips}, {Maluendes}, \&
  {Green}}]{Phillips1996}
{Phillips}, T.~R., {Maluendes}, S., \& {Green}, S. 1996, \apjs, 107, 467

\bibitem[{{Risacher} {et~al.}(2006){Risacher}, {Vassilev}, {Monje}, {Lapkin},
  {Belitsky}, {Pavolotsky}, {Pantaleev}, {Bergman}, {Ferm}, {Sundin},
  {Svensson}, {Fredrixon}, {Meledin}, {Gunnarsson}, {Hagstr{\"o}m},
  {Johansson}, {Olberg}, {Booth}, {Olofsson}, \& {Nyman}}]{Risacher_etal2006}
{Risacher}, C., {Vassilev}, V., {Monje}, R., {et~al.} 2006, \aap, 454, L17

\bibitem[{{Sch{\"o}ier} {et~al.}(2005){Sch{\"o}ier}, {van der Tak}, {van
  Dishoeck}, \& {Black}}]{Schoier2005}
{Sch{\"o}ier}, F.~L., {van der Tak}, F.~F.~S., {van Dishoeck}, E.~F., \&
  {Black}, J.~H. 2005, \aap, 432, 369

\bibitem[{{Snell} {et~al.}(2000{\natexlab{a}}){Snell}, {Howe}, {Ashby},
  {Bergin}, {Chin}, {Erickson}, {Goldsmith}, {Harwit}, {Kleiner}, {Koch},
  {Neufeld}, {Patten}, {Plume}, {Schieder}, {Stauffer}, {Tolls}, {Wang},
  {Winnewisser}, {Zhang}, \& {Melnick}}]{Snell2000a}
{Snell}, R.~L., {Howe}, J.~E., {Ashby}, M.~L.~N., {et~al.} 2000{\natexlab{a}},
  \apjl, 539, L97

\bibitem[{{Snell} {et~al.}(2000{\natexlab{b}}){Snell}, {Howe}, {Ashby},
  {Bergin}, {Chin}, {Erickson}, {Goldsmith}, {Harwit}, {Kleiner}, {Koch},
  {Neufeld}, {Patten}, {Plume}, {Schieder}, {Stauffer}, {Tolls}, {Wang},
  {Winnewisser}, {Zhang}, \& {Melnick}}]{Snell2000b}
{Snell}, R.~L., {Howe}, J.~E., {Ashby}, M.~L.~N., {et~al.} 2000{\natexlab{b}},
  \apjl, 539, L101

\bibitem[{{Spergel} {et~al.}(2007){Spergel}, {Bean}, {Dor{\'e}}, {Nolta},
  {Bennett}, {Dunkley}, {Hinshaw}, {Jarosik}, {Komatsu}, {Page}, {Peiris},
  {Verde}, {Halpern}, {Hill}, {Kogut}, {Limon}, {Meyer}, {Odegard}, {Tucker},
  {Weiland}, {Wollack}, \& {Wright}}]{Spergel2007}
{Spergel}, D.~N., {Bean}, R., {Dor{\'e}}, O., {et~al.} 2007, \apjs, 170, 377

\bibitem[{{Subrahmanyan} {et~al.}(1992){Subrahmanyan}, {Kesteven}, \& {Te
  Lintel Hekkert}}]{Subrahmanyan1992}
{Subrahmanyan}, R., {Kesteven}, M.~J., \& {Te Lintel Hekkert}, P. 1992, \mnras,
  259, 63

\bibitem[{{Subrahmanyan} {et~al.}(1990){Subrahmanyan}, {Narasimha},
  {Pramesh-Rao}, \& {Swarup}}]{Subrahmanyan1990}
{Subrahmanyan}, R., {Narasimha}, D., {Pramesh-Rao}, A., \& {Swarup}, G. 1990,
  \mnras, 246, 263

\bibitem[{{van der Tak} {et~al.}(2007){van der Tak}, {Black}, {Sch{\"o}ier},
  {Jansen}, \& {van Dishoeck}}]{VanderTak2007}
{van der Tak}, F.~F.~S., {Black}, J.~H., {Sch{\"o}ier}, F.~L., {Jansen}, D.~J.,
  \& {van Dishoeck}, E.~F. 2007, \aap, 468, 627

\bibitem[{{Wiklind} \& {Combes}(1996)}]{WiklindCombes1996}
{Wiklind}, T. \& {Combes}, F. 1996, \nat, 379, 139

\bibitem[{{Wiklind} \& {Combes}(1998)}]{WiklindCombes1998}
{Wiklind}, T. \& {Combes}, F. 1998, \apj, 500, 129

\bibitem[{{Wiklind} \& {Combes}(2005)}]{WiklindCombes2005}
{Wiklind}, T. \& {Combes}, F. 2005, in Bulletin of the American Astronomical
  Society, Vol.~37, Bulletin of the American Astronomical Society, 1424

\bibitem[{{Wilson} {et~al.}(2006){Wilson}, {Henkel}, \&
  {H{\"u}ttemeister}}]{Wilson2006}
{Wilson}, T.~L., {Henkel}, C., \& {H{\"u}ttemeister}, S. 2006, \aap, 460, 533

\bibitem[{{Winn} {et~al.}(2002){Winn}, {Kochanek}, {McLeod}, {Falco}, {Impey},
  \& {Rix}}]{Winn2002}
{Winn}, J.~N., {Kochanek}, C.~S., {McLeod}, B.~A., {et~al.} 2002, \apj, 575,
  103

\bibitem[{{Zeng} {et~al.}(1984){Zeng}, {Batrla}, \& {Wilson}}]{Zeng1984}
{Zeng}, Q., {Batrla}, W., \& {Wilson}, T.~L. 1984, \aap, 141, 127

\end{thebibliography}

\clearpage

\end{document}